\documentclass[twocolumn,trackchanges]{aastex631}
\received{}
\revised{}
\accepted{}

\submitjournal{ApJ}

\shorttitle{Magnetic Flux Emergence from the Solar Convection Zone}
\shortauthors{Toriumi et al.}
\graphicspath{{./}{figures/}}
\begin{document}

\title{Convective Magnetic Flux Emergence Simulations from the Deep Solar Interior to the Photosphere: Comprehensive Study of Flux Tube Twist}

\correspondingauthor{Shin Toriumi}
\email{toriumi.shin@jaxa.jp}

\author[0000-0002-1276-2403]{Shin Toriumi}
\affiliation{Institute of Space and Astronautical Science, Japan Aerospace Exploration Agency, 3-1-1 Yoshinodai, Chuo-ku, Sagamihara, Kanagawa 252-5210, Japan}

\author[0000-0002-6312-7944]{Hideyuki Hotta}
\author[0000-0002-6814-6810]{Kanya Kusano}
\affiliation{Institute for Space-Earth Environmental Research, Nagoya University, Furo-cho, Chikusa-ku, Nagoya 464-8601, Japan}



\begin{abstract}
The emergence of magnetic flux from the deep convection zone plays an important role in the solar magnetism, such as the generation of active regions and triggering of various eruptive phenomena, including jets, flares, and coronal mass ejections. To investigate the effects of magnetic twist on flux emergence, we performed numerical simulations of flux tube emergence using the radiative magnetohydrodynamic code R2D2, and conducted a systematic survey on the initial twist. Specifically, we varied the twist of the initial tube both positively and negatively from zero to twice the critical value for kink instability. As a result, regardless of the initial twist, the flux tube was lifted by the convective upflow and reached the photosphere to create sunspots. However, when the twist was too weak, the photospheric flux was quickly diffused and not retained long as coherent sunspots. The degree of magnetic twist measured in the photosphere conserved the original twist relatively well, and was comparable to actual solar observations. Even in the untwisted case, a finite amount of magnetic helicity was injected into the upper atmosphere because the background turbulence added helicity. However, when the initial twist exceeded the critical value for kink instability, the magnetic helicity normalized by the total magnetic flux was found to be unreasonably larger than the observations, indicating that the kink instability of the emerging flux tube may not be a likely scenario for the formation of flare-productive active regions.
\end{abstract}

\keywords{Magnetohydrodynamics (1964) --- Solar magnetic fields (1503) --- Solar interior (1500) --- Sunspots (1653) --- Solar flares (1496) --- Solar coronal mass ejections (310) --- Emerging flux tubes (458)}


\section{Introduction} \label{sec:intro}

In the generally-accepted scenario of the solar magnetism, sunspots and active regions (ARs) are formed by the emergence of a toroidal flux tube from the deep convection zone, which is amplified by the dynamo mechanism \citep[flux emergence:][]{1955ApJ...121..491P,1961ApJ...133..572B}. The spots inject magnetic helicity into the corona, and the excess magnetic energy (free energy) stored in the corona is released as solar flares through the close interplay between magnetic reconnection and magnetohydrodynamic (MHD) instability \citep{2002A&ARv..10..313P,2011SSRv..159...19F,2011LRSP....8....6S,2019LRSP...16....3T}. This indicates that the twisting of the emerging flux is one of the most critical physical parameters in the entire life cycle of the generation, transport, and release of a magnetic field \citep{2008JGRA..113.3S04A,2014LRSP...11....3C,2015LRSP...12....1V,2021LRSP...18....5F}.

For instance, it has been suggested that the flux tube in the interior must have sufficient twisting to maintain its integrity so that it can successfully reach the photosphere. Without a twist, the magnetic flux would be stripped away by the counteracting flow while the tube rises in the interior \citep{1979A&A....71...79S,1996ApJ...472L..53M,1998ApJ...492..804E} \citep[see Section 5.3 of][for a thorough discussion]{2021LRSP...18....5F}.

According to the series of three-dimensional (3D) flux emergence simulations, as pioneered by \citet{2001ApJ...554L.111F}, \citet{2003ApJ...586..630M}, and \citet{2004A&A...426.1047A}, the flux tube appearing in the photosphere forms a pair of flux concentrations of positive and negative polarities, and due to the tube's twist, these polarities show a yin-yang pattern called magnetic tongues \citep{2011SoPh..270...45L,2015SoPh..290..727P} in the photospheric surface. As the flux tube continues to emerge, the tilt angle of the two spots varies, whereas the spots themselves show rotational motion \citep{2015A&A...582A..76S}. These photospheric motions together inject free energy and magnetic helicity into the corona \citep{2003ApJ...586..630M}. Parameter surveys in which the twist strength of the initial subsurface flux tube was varied were conducted to study how the characteristics of the emerging flux and resultant AR depend on the twist \citep[for example,][]{2006A&A...460..909M,2011PASJ...63..407T,2013A&A...553A..55T,2016A&A...593A..63S}. These surveys commonly show that the rising speed, spot rotation, and free-energy injection into the corona are more pronounced for a stronger twist.

The long history of solar observations has revealed that ARs that produce massive flares are prone to complex morphologies \citep{2019LRSP...16....3T}. Of particular importance is that sunspots in which umbrae of the positive and negative polarities are closely adjoined and surrounded by a single penumbra, called $\delta$-spots, show outstanding flare activity \citep{1960AN....285..271K,2000ApJ...540..583S,2017ApJ...834...56T}. One of the formation processes suggested for $\delta$-spots is the kink instability of the emerging flux tube owing to its strong magnetic twist \citep{1991SoPh..136..133T,1996ApJ...469..954L,1999ApJ...522.1190L}. \citet{1998ApJ...505L..59F,1999ApJ...521..460F} performed 3D flux emergence simulations within the convection zone and showed that a flux tube with the initial twist strength exceeding the critical value for kink instability becomes deformed as it rises, which occurs as a result of the twist about the tube axis converting to the writhe of the axis itself. \citet{2015ApJ...813..112T} showed that when a kink-unstable flux tube reaches the photosphere, it displays two strongly rotating spots and complex magnetic field structures resembling the observed $\delta$-spots. \citet{2017ApJ...850...39T} also demonstrated that ARs formed by kink instability could inject the largest free energy of the four types of flux emergence processes suggested for flare-productive ARs \citep{2017ApJ...834...56T}. \citet{2018ApJ...864...89K} further extended the simulations with extreme twists of up to four times the critical value for instability, revealing that the magnetic structure became far more complicated. However, these numerical simulations did not account for the effects of the intense buffeting of external thermal convection; therefore, the actual results are unclear.

The AR formation by convective flux emergence was modeled by \citet{2010ApJ...720..233C} and \citet{2014ApJ...785...90R}. However, they introduced flux tubes kinematically from the bottom boundary located {\it only} at a depth of 15 Mm, which leaves the open question of how deep large-scale convection cells affect the flux tubes \citep[see also][]{2017ApJ...846..149C}. This drawback is overcome by the radiative MHD code R2D2, which can self-consistently reproduce thermal convection of various scales over the entire convection zone from the bottom ($-200$ Mm) to the photosphere \citep{2019SciA....5.2307H}. Using this code, \citet{2020MNRAS.494.2523H} performed flux emergence simulations and examined the magnetic and velocity field structures of the reproduced sunspots including penumbrae. Through the interaction of magnetic field and thermal convection, penumbral structures are naturally formed around the umbrae in the photosphere. \citet{2019ApJ...886L..21T} and \citet{2020MNRAS.498.2925H} modeled the spontaneous generation of $\delta$-spots, in which umbrae of opposite polarities are enclosed in a single penumbra, as a result of the collision of positive and negative polarities \citep[see also][]{2022MNRAS.517.2775K}.

One of the most recent findings from R2D2 flux emergence simulations was reported by \citet{2023NatSR..13.8994T}, who revealed that even when the initial flux tube is untwisted, it can reach the photosphere, aided by large-scale convective upflows, and inject finite (non-zero) magnetic helicity into the upper atmosphere. Detailed analysis showed that vortices developed just below the sunspots, which rotated the vertical magnetic fluxes to supply magnetic helicity to the atmosphere. While the emergence of untwisted flux tube was previously calculated by \citet{2021ApJ...907...19K} using semi-torus flux tubes \citep{2009A&A...503..999H} as the initial conditions, \citet{2023NatSR..13.8994T} demonstrated that an untwisted horizontal flux tube in a convective circumstances was capable of reaching the photosphere. Furthermore, using the flare prediction scheme proposed by \citet{2020Sci...369..587K}, they showed that the injected magnetic energy of the untwisted flux tube was sufficient to explain the medium-sized solar eruptions. These results present a new perspective of thermal convection as a non-negligible supplier of magnetic helicity to flare eruptions.

The purpose of this study is to extend the calculations in \citet{2023NatSR..13.8994T} to investigate the dependence of convective flux emergence on the magnetic twist. Specifically, we surveyed the characteristics of the generated ARs by varying the twist strength of the initial flux tube, including those exceeding the threshold value for kink instability, and provided theoretical constraints on the physical state of the subsurface emerging flux.

The remainder of this paper is organized as follows. We introduce the numerical setup in Section \ref{sec:setup}, and explain the analysis in Section \ref{sec:analysis}. Section \ref{sec:results} presents the results of the analysis. Finally, the results are summarized and discussed in Section \ref{sec:discussion}.

\section{Numerical setup}\label{sec:setup}

The basic numerical setup is consistent with that adopted by \citet{2023NatSR..13.8994T}. We used the radiative MHD code R2D2 \citep{2019SciA....5.2307H}, which solves the radiative transfer and equation of state by employing the reduced speed-of-sound technique (RSST) \citep{2012A&A...539A..30H,2015ApJ...798...51H,2019A&A...622A.157I}. The basic equations are as follows.
\begin{eqnarray}
  \frac{\partial\rho_{1}}{\partial t} &=& -\frac{1}{\xi^{2}}\nabla\cdot(\rho\mbox{\boldmath $V$}),\\
  \frac{\partial\rho_{1}}{\partial t} &=& -\nabla\cdot(\rho\mbox{\boldmath $V$}\mbox{\boldmath $V$})-\nabla p_{1}+\rho_{1}\mbox{\boldmath $g$}+\frac{1}{4\pi}(\nabla\times\mbox{\boldmath $B$})\times\mbox{\boldmath $B$},\\
  \frac{\partial \mbox{\boldmath $B$}}{\partial t} &=& \nabla\times(\mbox{\boldmath $V$}\times\mbox{\boldmath $B$}),\\
  \rho T\frac{\partial s_{1}}{\partial t} &=& \rho T(\mbox{\boldmath $V$}\cdot\nabla)s+Q,\\
  p_{1} &=& p_{1}(\rho, s), \label{eq:pressure}
\end{eqnarray}
where $\rho$, $\mbox{\boldmath $V$}$, $\mbox{\boldmath $B$}$, $p$, $T$, $s$, $\mbox{\boldmath $g$}$, $Q$, and $\xi$ are the density, velocity, magnetic field, gas pressure, temperature, entropy, gravitational acceleration in the vertical ($z$) direction, radiative heating, and the RSST factor, respectively. The subscript $1$ indicates the perturbation from the stationary stratification in the $z$-direction, which is indicated by the subscript $0$, namely,
\begin{eqnarray}
  \rho &=& \rho_{0}+\rho_{1},\\
  p &=& p_{0}+p_{1},\\
  s &=& s_{0}+s_{1}.
\end{eqnarray}
The background stratification was based on Model S \citep{1996Sci...272.1286C}, while in Equation (\ref{eq:pressure}), we used the equation of state considering the partial ionization effect following the OPAL repository \citep{1996ApJ...456..902R}. The radiative transfer equation was solved by employing the gray approximation and Rosseland mean opacity. In the present setup, only the upward and downward radiative transfers were considered. The RSST factor $\xi$ was introduced to suppress the fast sound speed in the convection zone so that the computation was accelerated:
\begin{eqnarray}
  \xi(z)=\max{\left(
    1,
    \xi_{0}\left[\frac{\rho_{0}(z)}{\rho_{\rm b}}\right]^{1/3},
    \frac{c_{\rm s}(z)}{c_{\rm b}}
    \right)},
\end{eqnarray}
where $\xi_{0}=160$ was adopted. $\rho_{\rm b}=0.2\ {\rm g}\ {\rm cm}^{-3}$ and $c_{\rm b}=2.2\times 10^{7}\ {\rm cm}\ {\rm s}^{-1}$ are the density and sound speed around the bottom of the convection zone, respectively, while $c_{\rm s}=\sqrt{(\partial p/\partial\rho)_{\rm s}}$ is the local adiabatic sound speed.

\begin{figure}
\begin{center}
\includegraphics[width=0.5\textwidth]{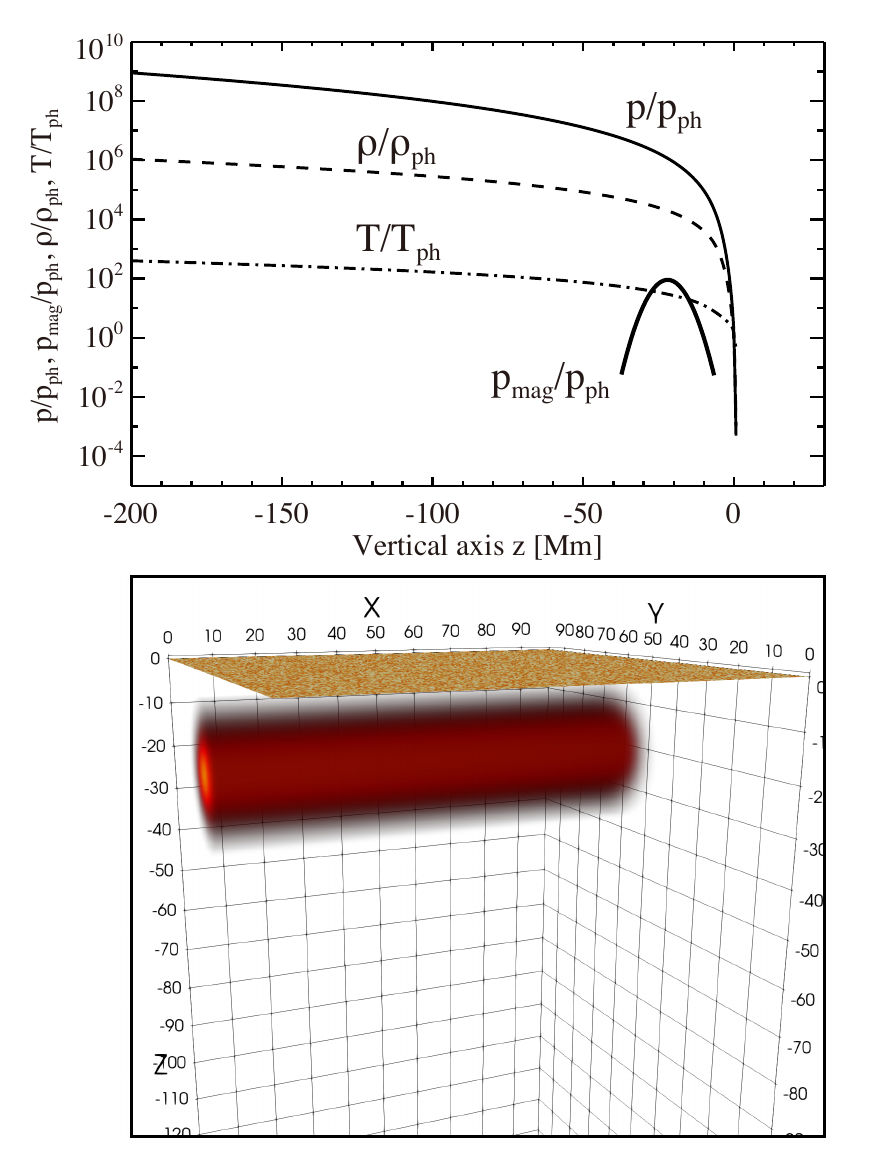}
\end{center}
\caption{(Top) Vertical profiles along the $z$-axis of the initial background gas pressure $p$ (thin solid), density $\rho$ (dashed), and temperature $T$ (dash-dotted), normalized by their photospheric values. Overplotted by the thick solid line is the magnetic pressure, $p_{\rm mag}=B^{2}/(8\pi)$, along the vertical axis crossing the flux tube center for the case of $q/q_{\rm cr}=0$. The typical values in the photospheric level are $p_{\rm ph}=6.6\times 10^{4}\ {\rm dyn}\ {\rm cm}^{-2}$, $\rho_{\rm ph}=1.8\times 10^{-7}\ {\rm g}\ {\rm cm}^{-3}$, $T_{\rm ph}=5500\ {\rm K}$. (Bottom) 3D volume rendering of the total magnetic field strength, $|\mbox{\boldmath $B$}|=(B_{x}^{2}+B_{y}^{2}+B_{z}^{2})^{1/2}$, for the $q/q_{\rm cr}=0$ case at $t=0\ {\rm hr}$. The 2D slice near the top presents the emergent intensity.\label{fig:initial}}
\end{figure}

We conducted the convection simulation without introducing magnetic fields in a 3D Cartesian domain with an extent of $(L_{x}, L_{y}, L_{z})=(98.3\ {\rm Mm}, 98.3\ {\rm Mm}, 201.7\ {\rm Mm})$, which was resolved by a $1024\times 1024\times 384$ grid, until the convection reached statistical equilibrium (see the top panel of Figure \ref{fig:initial}). The bottom boundary was located at $201\ {\rm Mm}$ below the mean $\tau=1$ surface (that is, the $z=0$ layer), whereas the top boundary was $700\ {\rm km}$ above the $\tau=1$ surface. The grid spacing in both horizontal directions was $\Delta x=\Delta y=96\ {\rm km}$ (uniform). The spacing for the vertical direction was $\Delta z=48\ {\rm km}$ from the top boundary to $z=-5.6\ {\rm Mm}$, and it linearly increased to $\Delta z=1486\ {\rm km}$ at the bottom boundary. A periodic boundary condition was applied in both the horizontal directions, whereas a potential field condition was used for the magnetic field at the top boundary.

Thermal convection was excited as a result of the energy input at the bottom boundary of the Cartesian box and the radiative transfer. The amount of injected energy at the bottom boundary was equal to the energy flux density in the solar photosphere. In the actual solar convection zone, due to the curvature effect, the energy flux density is radius-dependent and increases with depth, which is adjusted in the R2D2 code by controlling the radiation diffusion coefficient. As a side effect, the thermal convection velocity may differ from reality. However, the error is sufficiently small because the convection velocity only depends on the energy flux density to the power of $1/3$. In addition, it was confirmed that the energy flux is almost constant in the $z$-direction in the R2D2 simulations \citep[see Figure 3 of][]{2019SciA....5.2307H} because the loss of kinetic and magnetic energies dissipated by the artificial viscosity is transformed into an increase in internal energy.

We then introduced a horizontal magnetic flux tube oriented in the $x$-direction and defined this timing as $t=0$. The magnetic field in the initial flux tube has the following form:
\begin{eqnarray}
  B_{x}(r)=B_{\rm tb}\exp{\left(-\frac{r^{2}}{R_{\rm tb}^{2}}\right)},
  \label{eq:bx}
\end{eqnarray}
and
\begin{eqnarray}
  B_{\phi}(r)=qrB_{\rm tb}(r),
  \label{eq:bphi}
\end{eqnarray}
where $B_{\rm tb}$, $r$, $\phi$, $R_{\rm tb}$, and $q$ denote the axial field strength, radial distance from the tube axis $(y_{\rm tb}, z_{\rm tb})$, i.e., $r=[(y-y_{\rm tb})^2+(z-z_{\rm tb})^{2}]^{1/2}$, azimuth direction about the tube axis in the $y$--$z$ plane, typical tube radius, and the twist strength, respectively. This type of Gaussian flux tube has been used in many previous flux emergence simulations referred to in Section \ref{sec:intro}. The center of the tube was placed at $(y_{\rm tb}, z_{\rm tb})=(61.5\ {\rm Mm}, -22.0\ {\rm Mm})$, while the typical radius was set to be $R_{\rm tb}=8\ {\rm Mm}$. (see Figure \ref{fig:initial}).

The entropy inside the flux tube was adjusted so that the internal gas pressure was enhanced by
\begin{eqnarray}
  \delta p_{\rm exc}=\frac{B_{x}^{2}(r)}{8\pi}
  \left[
    q^{2} \left( \frac{R_{\rm tb}^{2}}{2} \right)-1
    \right]\,(<0)
\end{eqnarray}
whereas the density was kept the same, which allowed the tube to have no magnetic buoyancy at $t=0$. In other words, the tube was advected by external flows.

We varied parameter $q$ to investigate the effect of the initial twist. In the remainder of this paper, we use the twist strength normalized by its critical value for kink instability. According to \citet{1996ApJ...469..954L}, the critical twist for kink instability for a Gaussian flux tube is given by $q_{\rm cr}=1/R_{\rm tb}=0.125\ {\rm Mm}^{-1}$ (at the beginning of the simulation at $t=0$). We investigated nine simulation cases, in which $q/q_{\rm cr}=[-2, -1, -1/2, -1/4, 0, 1/4, 1/2, 1, 2]$. A positive (negative) value of $q/q_{\rm cr}$ indicates a right-handed (left-handed) twist, and the cases with $|q/q_{\rm cr}|\geq 1$ are kink-unstable.

However, increasing the flux tube twist while maintaining the axial field $B_{x}$ enhances the azimuthal field $B_{\phi}$, and thus, the total magnetic energy,
\begin{eqnarray}
  E_{\rm mag}=\int\frac{B^{2}}{8\pi}\,dV,
\end{eqnarray}
differs in the simulation cases. Therefore, for a fair comparison, we also controlled the axial field strength at the tube center $B_{\rm tb}$ such that the magnetic energies $E_{\rm mag}$ were the same throughout the models.

Table \ref{tab:cases} summarizes the simulation cases and parameters of the initial flux tubes, including the magnetic flux within the $(y, z)$ cross-section:
\begin{eqnarray}
  \Phi_{x}=\int B_{x}\, dS.
\end{eqnarray}

\begin{deluxetable*}{cDDcDc}
\tablecaption{Simulation cases and flux-tube parameters\label{tab:cases}}
\tablewidth{0pt}
\tablehead{
\colhead{Case} & \multicolumn2c{$B_{\rm tb}$} & \multicolumn2c{$R_{\rm tb}$} & \colhead{$q/q_{\rm cr}$} & \multicolumn2c{$q$} & \colhead{$\Phi_{x}$} \\
\nocolhead{} & \multicolumn2c{(kG)} & \multicolumn2c{(Mm)} & \nocolhead{} & \multicolumn2c{(${\rm Mm}^{-1}$)} & \colhead{(Mx)}
}
\decimalcolnumbers
\startdata
1 & $7.1$ & $8.0$ & $-2$ & $-0.25$ & $1.40\times 10^{22}$ \\
2 & $10.0$ & $8.0$ & $-1$ & $-0.125$ & $1.97\times 10^{22}$ \\
3 & $11.5$ & $8.0$ & $-1/2$ & $-0.0625$ & $2.28\times 10^{22}$ \\ 
4 & $12.1$ & $8.0$ & $-1/4$ & $-0.03125$ & $2.38\times 10^{22}$ \\
5 & $12.2$ & $8.0$ & $0$ & $0$ & $2.42\times 10^{22}$ \\
6 & $12.1$ & $8.0$ & $1/4$ & $0.03125$ & $2.38\times 10^{22}$ \\
7 & $11.5$ & $8.0$ & $1/2$ & $0.0625$ & $2.28\times 10^{22}$ \\
8 & $10.0$ & $8.0$ & $1$ & $0.125$ & $1.97\times 10^{22}$ \\
9 & $7.1$ & $8.0$ & $2$ & $0.25$ & $1.40\times 10^{22}$ \\
\enddata
\tablecomments{The total magnetic energy is set to be $5.85\times 10^{34}\ {\rm erg}$ for all cases. Cases 5, 6, and 7 correspond to the non-twisted, weakly-twisted, and strongly-twisted cases, respectively, in \citet{2023NatSR..13.8994T}.}
\end{deluxetable*}

\section{Analysis}\label{sec:analysis}

The magnetic and velocity fields on the photospheric surface were measured. However, because the velocity at the $z=0$ layer (i.e., the mean $\tau=1$ layer) was perturbed by strong downflows, the quantities were measured at $z=200\ {\rm km}$,\footnote{200 km is an approximate pressure scale height near the photosphere.} where the velocity structure was somewhat relaxed. We applied 6-hr moving averaging to all time-variation data.\footnote{The 6-hr smoothing window was chosen because it was longer than the time scales of convective fluctuations (a few 10 min to a few hr) but still shorter than the typical emergence duration (30 to 40 hr).}

The total unsigned magnetic flux in the photosphere is defined as
\begin{eqnarray}
  \Phi=\int_{S} |B_{z}|\, dS,
\end{eqnarray}
whereas the flux growth rate was obtained by taking its time derivative, $d\Phi/dt$. The sunspot area, $A_{\rm spot}$, was measured as the area where the emergent intensity was less than the 90\% of the quiet-Sun average, $I<0.9I_{0}$. The spot area is typically expressed in millionths of the solar hemisphere (MSH), which is equivalent to $3.0\times 10^{6}\ {\rm km}^{2}$.

To quantify the degree of magnetic field twisting, the twist parameter $\alpha$ is often used, which is the ratio between the electric current density $\mbox{\boldmath $J$}=\nabla\times\mbox{\boldmath $B$}$ and magnetic field $\mbox{\boldmath $B$}$, or, $\nabla\times\mbox{\boldmath $B$}=\alpha\mbox{\boldmath $B$}$. It is assumed here that the vectors $\nabla\times\mbox{\boldmath $B$}$ and $\mbox{\boldmath $B$}$ are parallel. From each photospheric magnetogram, we calculated the twist parameter $\alpha$ averaged over the entire horizontal extent \citep{1998ApJ...507..417L}:
\begin{eqnarray}
  \alpha_{\rm av}^{(0)}=\Big\langle \frac{J_{z}}{B_{z}} \Big\rangle.
  \label{eq:av}
\end{eqnarray}
We also considered two forms in which the weighting function was changed \citep{2004PASJ...56..831H}, namely,
\begin{eqnarray}
  \alpha_{\rm av}^{(1)}=\frac{\langle J_{z}\, {\rm sgn}(B_{z})\rangle}{\langle|B_{z}|\rangle}
  \label{eq:avabs}
\end{eqnarray}
and
\begin{eqnarray}
  \alpha_{\rm av}^{(2)}=\frac{\langle B_{z}J_{z}\rangle}{\langle B_{z}^{2}\rangle}.
  \label{eq:avsqr}
\end{eqnarray}
Although $\alpha_{\rm av}^{(0)}$ through $\alpha_{\rm av}^{(2)}$ are all parameters in which $\langle J_{z}\rangle$ is normalized by $\langle B_{z}\rangle$, there is a difference in the way the weighting by $B_{z}$ is given. Owing to this difference, $\alpha_{\rm av}^{(2)}$ is the most sensitive to strong $B_{z}$ regions. To measure $\alpha_{\rm av}$, we used only grid points where $|B_{z}|\ge 100\ {\rm G}$.

As the flux tube emerges, magnetic helicity,
\begin{eqnarray}
  H_{\rm R}=\int \left(\mbox{\boldmath $A$}+\mbox{\boldmath $A$}_{\rm p}\right)
  \cdot \left(\mbox{\boldmath $B$}-\mbox{\boldmath $B$}_{\rm p}\right)\, dV,
  \label{eq:relhel}
\end{eqnarray}
is injected into the atmosphere through the photosphere \citep{1984JFM...147..133B,1985CoPPC...9..111F}, where $\mbox{\boldmath $A$}$ is the vector potential of the magnetic field $\mbox{\boldmath $B$}$ (that is, $\mbox{\boldmath $B$}=\nabla\times\mbox{\boldmath $A$}$) and $\mbox{\boldmath $B$}_{\rm p}$ denotes the potential magnetic field.
Following \citet{1984JFM...147..133B}, the time derivative of Equation (\ref{eq:relhel}) is expressed as
\begin{eqnarray}
  \frac{dH_{\rm R}}{dt}
  = && -2\int \mbox{\boldmath $E$}\cdot\mbox{\boldmath $B$}\, dV \nonumber \\
  && +2\int_{S}
  \left[
    \left(\mbox{\boldmath $A$}_{\rm p}\cdot \mbox{\boldmath $B$}_{\rm h}\right) V_{z}
    -\left(\mbox{\boldmath $A$}_{\rm p}\cdot \mbox{\boldmath $V$}_{\rm h}\right) B_{z}
    \right]\, dS \nonumber \\
  && -2\int_{S} \frac{\partial \Psi}{\partial t} \mbox{\boldmath $A$}_{\rm p}
  \cdot \mbox{\boldmath $n$}\, dS,
  \label{eq:helflux1}
\end{eqnarray}
where $\Psi$ is the solution of $\nabla^{2}\Psi=0$ (satisfying $\mbox{\boldmath $B$}_{\rm p}=\nabla\Psi$). The first term of the right-hand side vanishes assuming the ideal MHD and the third term vanishes if used for the planer boundaries, both of which are true for the present simulations. Therefore, the reduced form of Equation (\ref{eq:helflux1}), i.e., the magnetic helicity flux for the ideal MHD, which we measured in the photosphere, is
\begin{eqnarray}
  %% \frac{dH_{\rm R}}{dt}=2\int_{S}
  %% \left[
  %%   \left(\mbox{\boldmath $A$}_{\rm p}\cdot \mbox{\boldmath $B$}_{\rm h}\right) V_{z}
  %%   -\left(\mbox{\boldmath $A$}_{\rm p}\cdot \mbox{\boldmath $V$}_{\rm h}\right) B_{z}
  %%   \right]\, dS.
  F_{z}=2\int_{S}
  \left[
    \left(\mbox{\boldmath $A$}_{\rm p}\cdot \mbox{\boldmath $B$}_{\rm h}\right) V_{z}
    -\left(\mbox{\boldmath $A$}_{\rm p}\cdot \mbox{\boldmath $V$}_{\rm h}\right) B_{z}
    \right]\, dS.
  \label{eq:helflux}
\end{eqnarray}
%%in the photosphere, and i
Integrating it over time, we obtained the net injected magnetic helicity:
\begin{eqnarray}
%%  H_{\rm R}=\int_{0}^{t} \frac{dH_{\rm R}}{dt'}\, dt'.
  \Delta H_{\rm R}=\int_{0}^{t} F_{z}\, dt'.
\end{eqnarray}
Although the helicity conservation may not be strictly satisfied due to numerical dissipation, top boundary condition, etc., the above proxy was used to estimate the amount of total injected helicity.
It has been suggested that injection of magnetic helicity (i.e., time variations of $F_{z}$ and $\Delta H_{\rm R}$) has a close relation with the occurrence of flares \citep[e.g.,][]{2002ApJ...574.1066M,2002ApJ...580..528M,2002ApJ...577..501K,2012ApJ...760...31K,2010ApJ...718...43P,2012ApJ...750...48P,2012ApJ...752L...9J,2019ApJ...887...64T}. To calculate the vector potential $\mbox{\boldmath $A$}_{\rm p}$, we adopt the method proposed by \citet{2001ApJ...560L..95C}.

To estimate the injection of magnetic energy into the atmosphere, we measured the Poynting flux across the photosphere:
\begin{eqnarray}
  S_{z}=\frac{1}{4\pi} \int_{S}
  \left[
    B_{\rm h}^{2} V_{z}
    -(\mbox{\boldmath $B$}_{\rm h}\cdot\mbox{\boldmath $V$}_{\rm h})B_{z}
    \right]\, dS.
\end{eqnarray}
The injected magnetic energy can be obtained by integrating this over time:
\begin{eqnarray}
  E_{\rm mag}=\int_{0}^{t} S_{z}\,dt'.
\end{eqnarray}

\section{Results}\label{sec:results}

\subsection{Overall evolution}

\begin{figure*}
%\begin{interactive}{animation}{fig2_ql.mp4}
\begin{center}
\includegraphics[width=0.9\textwidth]{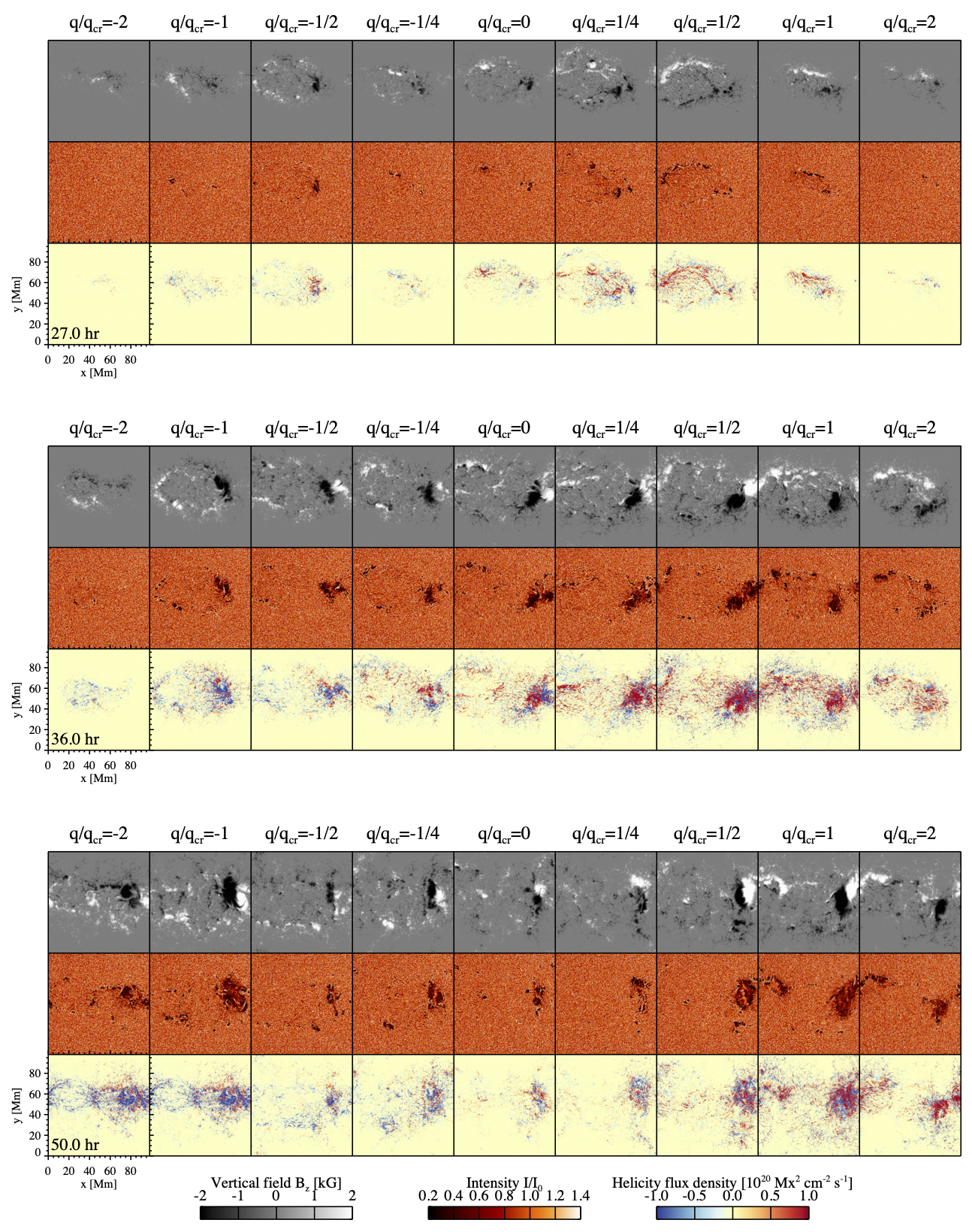}
\end{center}
%\end{interactive}
\caption{The vertical field strength, emergent intensity, and helicity flux density for the nine simulation cases with $q/q_{\rm cr}=-2$ to $2$. Shown are the time steps at $t=27.0\ {\rm hr}$, $36.0\ {\rm hr}$, and $50.0\ {\rm hr}$. The animated version of this figure shows the time evolutions from $t=0.0\ {\rm hr}$ to $120.0\ {\rm hr}$. The video duration is 16 seconds.\label{fig:ql}}
\end{figure*}

\begin{figure*}
%\begin{interactive}{animation}{fig3_3d.mp4}
\begin{center}
\includegraphics[width=0.7\textwidth]{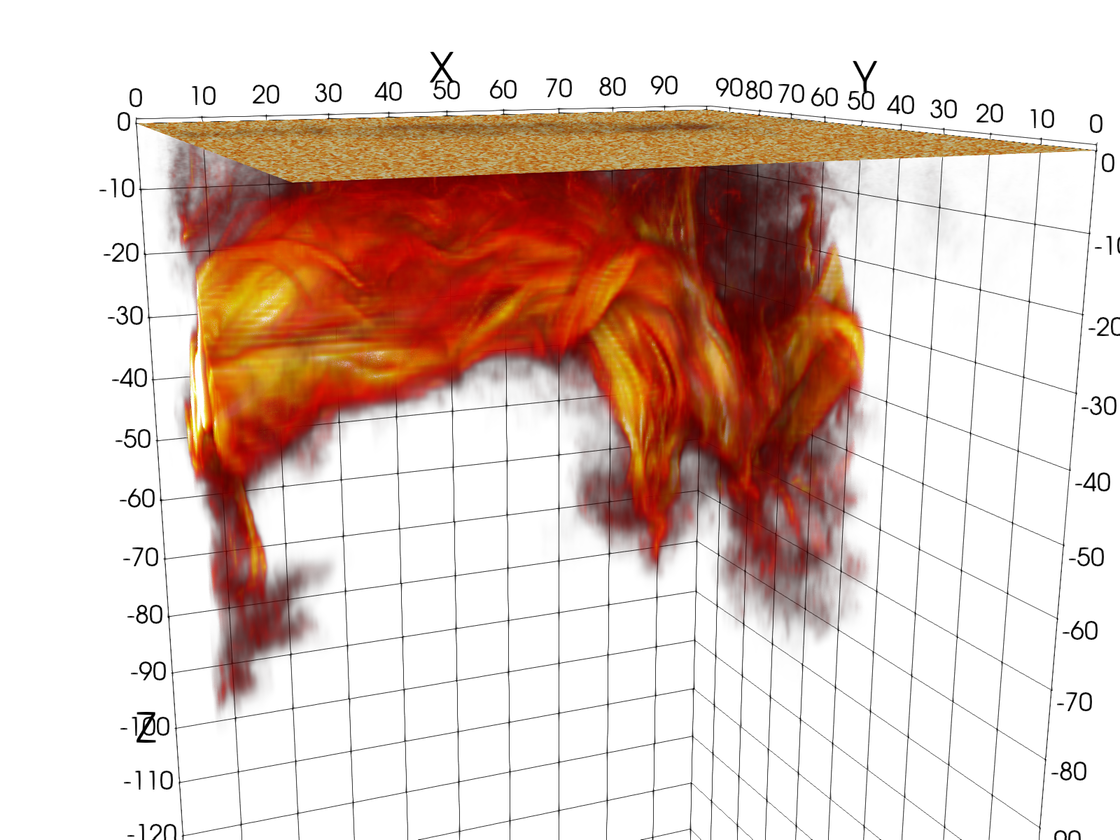}
\end{center}
%\end{interactive}
\caption{3D rendering of the total magnetic field strength at $t=24.0\ {\rm hr}$ for the case of $q/q_{\rm cr}=0$ is shown with the emergent intensity map in 2D slice near the top. The animated version of this figure shows the time evolutions from $t=0.0\ {\rm hr}$ to $60.0\ {\rm hr}$. The video duration is 8 seconds.\label{fig:3d}}
\end{figure*}

Figure \ref{fig:ql} and the corresponding animation show the temporal evolutions in the photosphere for the nine simulation cases, in which the vertical field strength, $B_{z}$, emergent intensity, $I/I_{0}$, and helicity flux density, $2\left[\left(\mbox{\boldmath $A$}_{\rm p}\cdot \mbox{\boldmath $B$}_{\rm h}\right) V_{z}-\left(\mbox{\boldmath $A$}_{\rm p}\cdot \mbox{\boldmath $V$}_{\rm h}\right) B_{z}\right]$, namely, the integrand of Equation (\ref{eq:helflux}), are plotted. The 3D view of the total magnetic field strength, $|\mbox{\boldmath $B$}|=(B_{x}^{2}+B_{y}^{2}+B_{z}^{2})^{1/2}$, and emergent intensity map for the case of $q/q_{\rm cr}=0$ are shown in Figure \ref{fig:3d}. The animated version provides the temporal evolution.

In the initial phase, weak magnetic fields are scattered in the photosphere owing to the turbulent dispersal of the rising flux tube. Soon after, from approximately $t=20\ {\rm hr}$, as the large-scale upflow pushes up the flux tube, the main body appears in the photosphere as a pair of positive and negative polarities, which fill the entire domain with a yin-yang pattern. Like in previous simulations in Section \ref{sec:intro}, at first, the direction of the separation, or the sunspot tilt, is not necessarily aligned in the $x$-direction (see snapshots at $t=27\ {\rm hr}$), but is gradually shifted toward it (for example, $t=34\ {\rm hr}$).

It should be noted that even in the untwisted case ($q/q_{\rm cr}=0$), the tilt angle clearly deviates from the $x$-direction owing to the strong influence of turbulence on the emerging flux in the convection zone \citep{2023NatSR..13.8994T}. From this result, we learn that it is incorrect to interpret just by seeing a yin-yang pattern in the photosphere that the emerging flux is twisted in the convection zone.

In all cases, from approximately $t=30\ {\rm hr}$, owing to the periodic boundary condition, the positive spot escapes from the left boundary ($x=0$) and returns to the domain from the right boundary ($x=98.3\ {\rm Mm}$). The positive spot collided with the negative spot and eventually formed a $\delta$-spot (around $t=50\ {\rm hr}$). The collision angle of both polarities and degree of spot rotation depend on the tube twist.

Injection of magnetic helicity occurs mainly around the spots, although the injection is apparently stronger in the penumbrae and polarity inversion lines between the colliding spots than that in the umbrae. The sign of magnetic helicity injection appears to be dominated by positive (red) for cases with $q/q_{\rm cr}>0$ and negative (blue) for $q/q_{\rm cr}<0$.

Another factor that is dependent on the twist is the spot decay. The magnetic flux for the untwisted case ($q/q_{\rm cr}=0$) is almost completely diffused by $t=70\ {\rm hr}$, whereas cases with stronger twists maintain their coherency, indicating that the flux tube twists protect the flux in the spots from being stripped away by the surrounding turbulence. It should also be noted that the cases with the strongest twists ($q/q_{\rm cr}=\pm 2$) have more scattered magnetic fields than that of the other cases.

The properties of the emerging fluxes and their dependence on the flux tube twist are further investigated in the following subsections.

\subsection{Magnetic flux and sunspot area}\label{subsec:flux}

\begin{figure*}
\begin{center}
\includegraphics[width=0.9\textwidth]{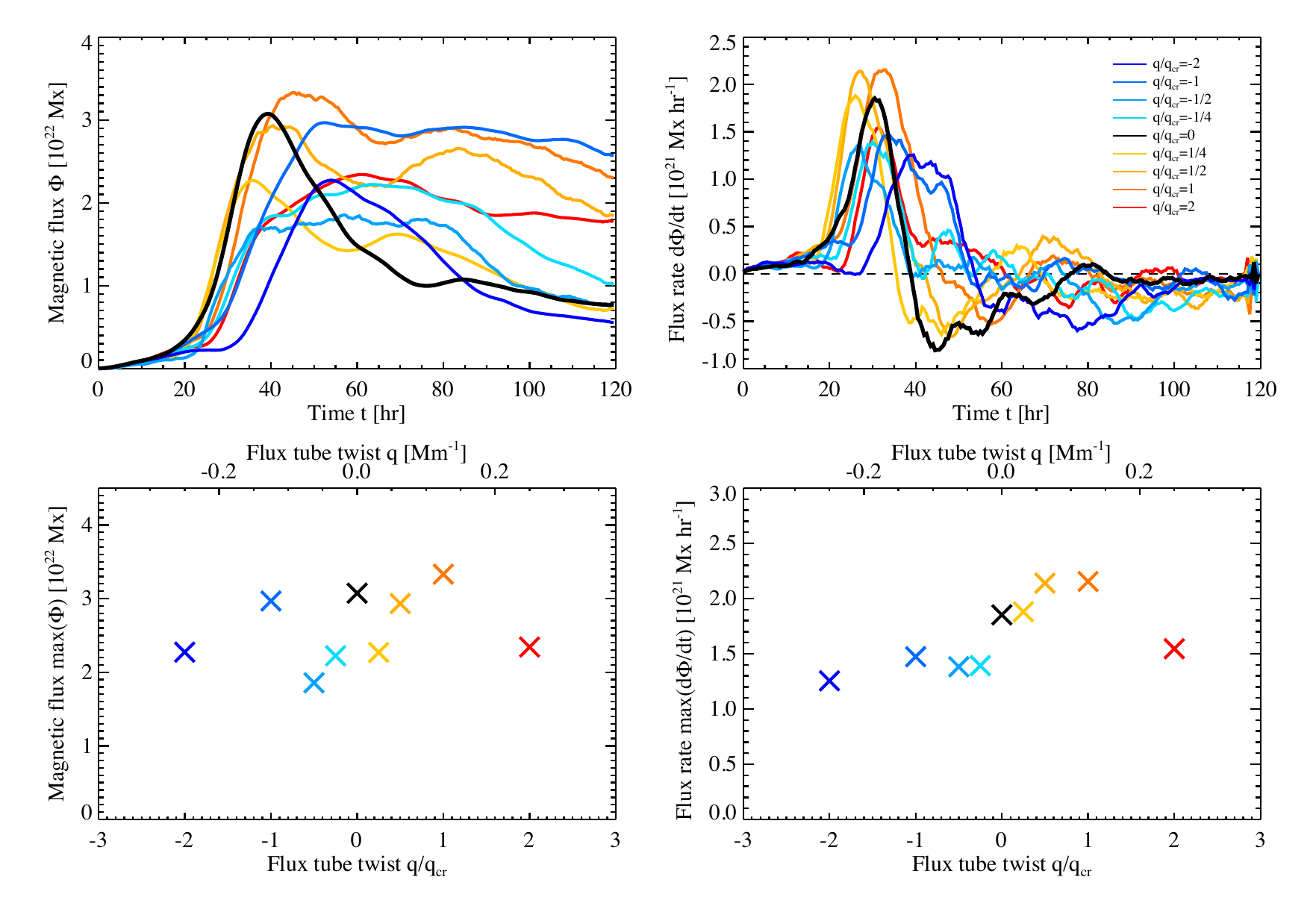}
\end{center}
\caption{(Top) Temporal evolutions of the total magnetic flux, $\Phi$, and the flux growth rate, $d\Phi/dt$, in the photosphere. (Bottom) Their peak values as a function of the initial flux tube twist, $q/q_{\rm cr}$.\label{fig:phi}}
\end{figure*}

The top panels of Figure \ref{fig:phi} show the temporal evolution of the total unsigned flux $\Phi$ and flux growth rate $d\Phi/dt$ in the photosphere. It can be seen that the emergence occurs in two stages. That is, during $t<20\ {\rm hr}$, the small-scale turbulence around the initial flux tube strips the surface of the tube and, thus, these fragmentary fluxes appear on the photosphere before the main body of the flux tube emerges. Then, after $t=20\ {\rm hr}$, the total flux breaks out, and the flux growth rate peaks, indicating that the main body of the flux tube appears on the photosphere.

In the bottom panels of Figure \ref{fig:phi}, the maximum surface fluxes $\max{(\Phi)}$ differ by a factor of two between the simulation cases. Contrary to the expectation that $\max{(\Phi)}$ increases with the twist, the cases with the largest twists ($q/q_{\rm cr}=\pm 2$) do not show large values of $\max{(\Phi)}$, and in particular, the case of $q/q_{\rm cr}=-2$ has the smallest flux growth rate $\max{(d\Phi/dt)}$, as shown in Figure \ref{fig:initial}. This can be attributed to the weak field strength in these cases. To ensure that the magnetic energies were the same for all cases, the initial axial field for these two cases was set to be $B_{\rm tb}=7.1\ {\rm kG}$. However, the equipartition field strength at this depth ($z_{\rm tb}=-22\ {\rm Mm}$) was on average $B_{\rm eq}=6.5\ {\rm kG}$, where $B_{\rm eq}^{2}/(8\pi)=\rho V^{2}/2$. Given that we used Gaussian-type flux tubes, as in Equations (\ref{eq:bx}) and (\ref{eq:bphi}), we expect that the azimuthal component in these cases will help maintain the coherency of the tube. However, because $B_{\rm tb}$ and thus the field strength of the entire flux tubes are weak, approximately comparable to $B_{\rm eq}$, they are easily collapsed by the external turbulent flows, and as a result, yield less pronounced magnetic fields in the photosphere.\footnote{To verify this hypothesis, we also tested two extreme cases where the twists were $q/q_{\rm cr}=\pm 4$ and the corresponding axial field strength was $B_{\rm tb}=4.1\ {\rm kG}$, and found that the total magnetic flux in the photosphere, $\max{(\Phi)}$, was even smaller.}

The magnetic flux curve peaks after $t=30\ {\rm hr}$ and then decays. However, there is diversity in the flux evolution, such as cases where the peak lasts for a long time and cases where it decreases rapidly.

\begin{figure}
\begin{center}
\includegraphics[width=0.5\textwidth]{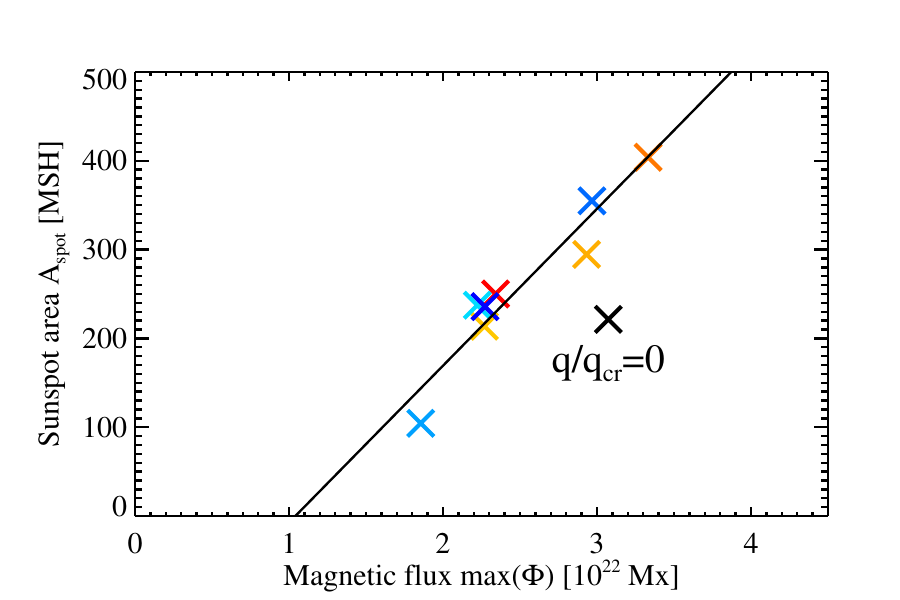}
\end{center}
\caption{Maximum photospheric magnetic flux, $\max{(\Phi)}$, and the corresponding sunspot area, $A_{\rm spot}$, for the nine cases. The straight line is the linear fit to the eight data points except for the untwisted flux tube, $q/q_{\rm cr}=0$, which is located much below the linear proportionality line.\label{fig:usflux_spotarea}}
\end{figure}

Figure \ref{fig:usflux_spotarea} compares the maximum magnetic flux, $\max{(\Phi)}$, and the sunspot area, $A_{\rm spot}$, at the time of $\max{(\Phi)}$, for the nine cases. In most cases, the two parameters show a linear proportionality indicated by a straight line. However, it is clearly seen that the untwisted case ($q/q_{\rm cr}=0$) falls significantly below the proportionality line, which indicates the tendency that an emerging AR with a weaker twist is more scattered, and thus, has a smaller amount of magnetic flux within the spots.

\begin{figure}
\begin{center}
\includegraphics[width=0.5\textwidth]{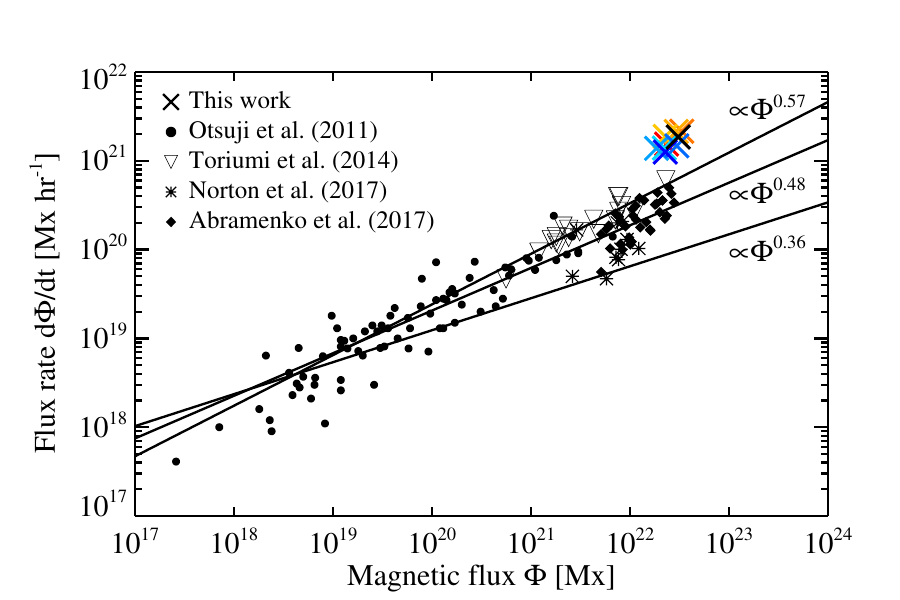}
\end{center}
\caption{Flux growth rate, $d\Phi/dt$, {\it versus} total magnetic flux, $\Phi$, for various observations and the present nine simulation cases. The observed values are adopted from \citet{2011PASJ...63.1047O}, \citet{2014ApJ...794...19T}, \citet{2017ApJ...842....3N}, and \citet{2017SoPh..292...48A}. The straight lines are the power-law fits to the observations: $d\Phi/dt\propto \Phi^{0.57}$ \citep{2011PASJ...63.1047O}, $d\Phi/dt\propto \Phi^{0.36}$ \citep{2017ApJ...842....3N}, and $d\Phi/dt\propto \Phi^{0.48}$ \citep{2019MNRAS.484.4393K}.\label{fig:phi_dphidt}}
\end{figure}

To examine if the simulation results can be justified, Figure \ref{fig:phi_dphidt} compares the total flux, $\Phi$, and the flux growth rate, $d\Phi/dt$, for the present simulation cases and the actual observations \citep{2011PASJ...63.1047O,2014ApJ...794...19T,2017ApJ...842....3N,2017SoPh..292...48A,2019MNRAS.484.4393K}. \citet{2017ApJ...842....3N} showed that numerical models tend to exhibit higher flux growth rates than observed values. Figure \ref{fig:phi_dphidt} indicates that this is also true for the present simulation results \citep[see also][]{2022MNRAS.517.2775K}. However, \citet{2017RNAAS...1...24S} reported the flux growth rate of $4.93\times 10^{20}\ {\rm Mx\ hr}^{-1}$ with a maximum of $1.12\times 10^{21}\ {\rm Mx\ hr}^{-1}$ for a total flux of $6.08\times 10^{22}\ {\rm Mx}$ in NOAA AR 12673, which does not differ significantly from our simulation results, considering that the plotted results are the peak values, $\max{(\Phi)}$ and $\max{(d\Phi/dt)}$. Therefore, we conclude that the present simulations adequately represent the Sun.

\subsection{Magnetic twist}\label{subsec:alpha}

\begin{figure*}
\begin{center}
\includegraphics[width=0.9\textwidth]{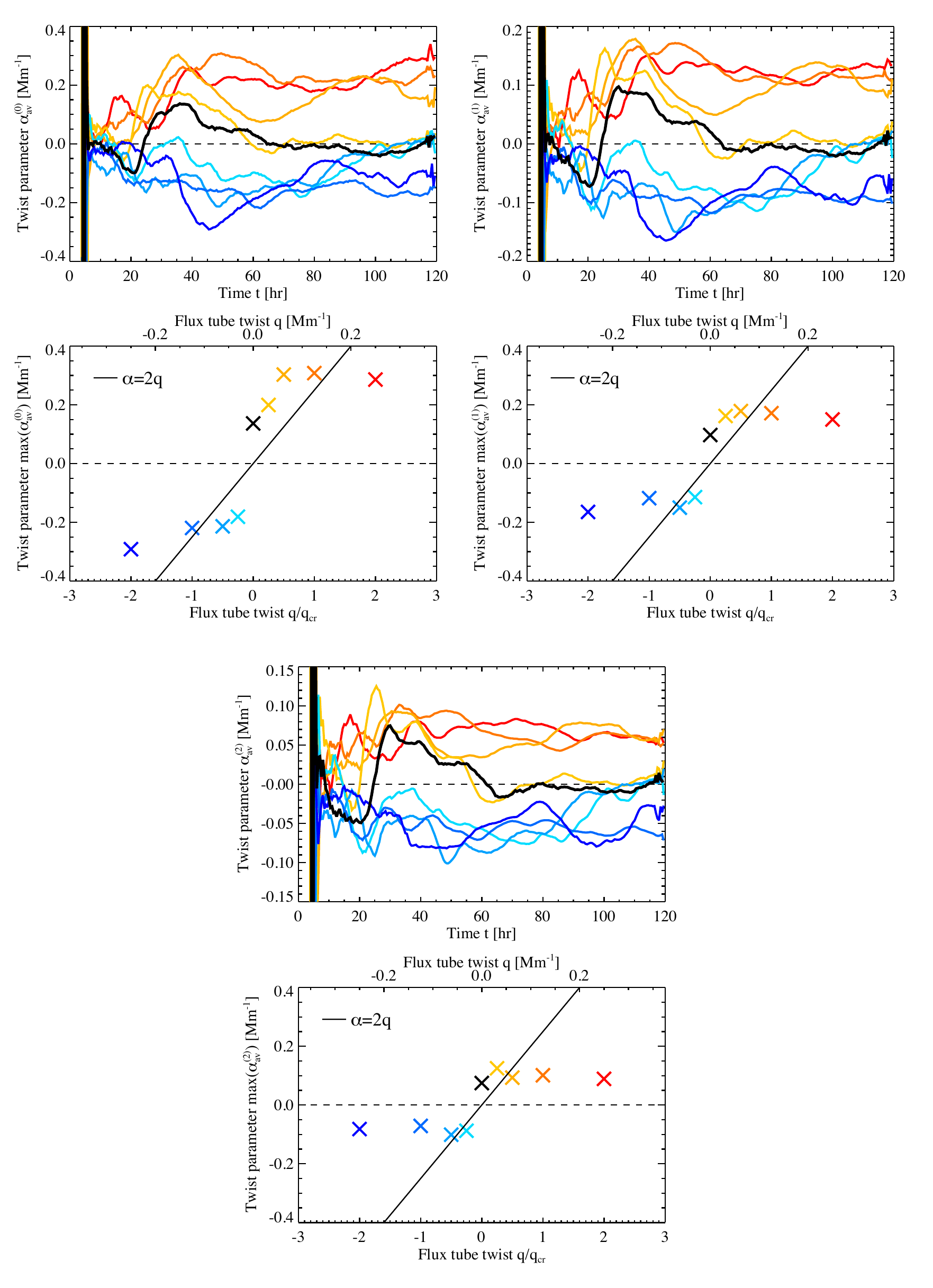}
\end{center}
\caption{Temporal evolutions of the twist parameters, $\alpha_{\rm av}^{(0)}$, $\alpha_{\rm av}^{(1)}$, and $\alpha_{\rm av}^{(2)}$, and their peak values measured between $t=10\ {\rm hr}$ and $110\ {\rm hr}$ as a function of the initial flux tube twist, $q/q_{\rm cr}$. For the negative twist cases ($q/q_{\rm cr}<0$), the peak values indicate the greatest negative values in the curves. The straight lines show the theoretical relation for the uniformly twisted flux tubes, $\alpha=2q$ \citep{1997ApJ...488..443L}.\label{fig:alpha}}
\end{figure*}

To examine how much of the twist in the initial flux tube was successfully transported to the photosphere by flux emergence, in Figure \ref{fig:alpha}, we plotted the temporal variations of the three twist parameters, $\alpha_{\rm av}^{(0)}$, $\alpha_{\rm av}^{(1)}$, and $\alpha_{\rm av}^{(2)}$ (Equations (\ref{eq:av})--(\ref{eq:avsqr})). The peak values are also shown as the function of the initial flux tube twist, $q/q_{\rm cr}$. We observe that $\max{(\alpha_{\rm av})}$ tends to decrease from $\alpha_{\rm av}^{(0)}$ to $\alpha_{\rm av}^{(2)}$. Considering that $\alpha_{\rm av}^{(2)}$ puts the largest weight to the strong-field pixels, this tendency may indicate that the weak-field pixels have a relatively large amount of magnetic twist.

Contrary to the previous study by \citet{2018ApJ...864...89K}, who reported that they could not find any clear dependence on the initial twist, Figure \ref{fig:alpha} shows that for each type of parameter, the peak value of $\alpha_{\rm av}$ exhibits a profound dependence on $q$. For a uniformly twisted flux tube, such as those adopted in this study, the relationship between the twist parameter $\alpha$ and the flux tube twist $q$ is given by $\alpha=2q$ \citep{1997ApJ...488..443L}, as indicated by the solid straight lines. For a smaller $|q|$, the measured $\alpha_{\rm av}$ roughly follows this theoretical line, which means that the flux tubes conserve the original twists throughout the emergence. However, $\alpha_{\rm av}$ becomes saturated for higher values of $|q|$ and deviates from the theoretical line. This could be attributed to the less successful emergence of strong-twist cases (see Section \ref{subsec:flux}), or may be because the background turbulence stripped away the magnetic twist from the flux tube during emergence. An alternative possibility is that, because the upper boundary of the computational box, which assumes the potential field condition, is close and only 500 km above, it mitigates the magnetic twist that was transported from the subsurface domain.

Measurements of $\alpha_{\rm av}$ on actual solar ARs have shown that $\alpha_{\rm av}$ is typically in the order of $0.01$ to $1\ {\rm Mm}^{-1}$ \citep[for example,][]{1996ApJ...462..547L,2004PASJ...56..831H,2010MNRAS.402L..30Z,2015PASJ...67....6O,2019MNRAS.484.4393K,2023ApJ...954L..20L}. Although we found in our simulations that the $\alpha_{\rm av}$ values can vary by a factor (but by less than one order) depending on the choice of the weighting function, the obtained $\alpha_{\rm av}$ values are within the range of the actual observations, regardless of the choice.

\subsection{Magnetic helicity}\label{subsec:helicity}

In \citet{2023NatSR..13.8994T}, we compared the cases of $q/q_{\rm cr}=1/2$, $1/4$, and $0$ and found that all cases injected positive magnetic helicity, even for the untwisted case. The background convection was identical to that in the present model, in which solar rotation was not considered, and it was coincidental that the sign of the injected helicity was positive for the three cases. What will then be the sign of magnetic helicity when the twist of the initial tube is negative?

\begin{figure*}
\begin{center}
\includegraphics[width=0.9\textwidth]{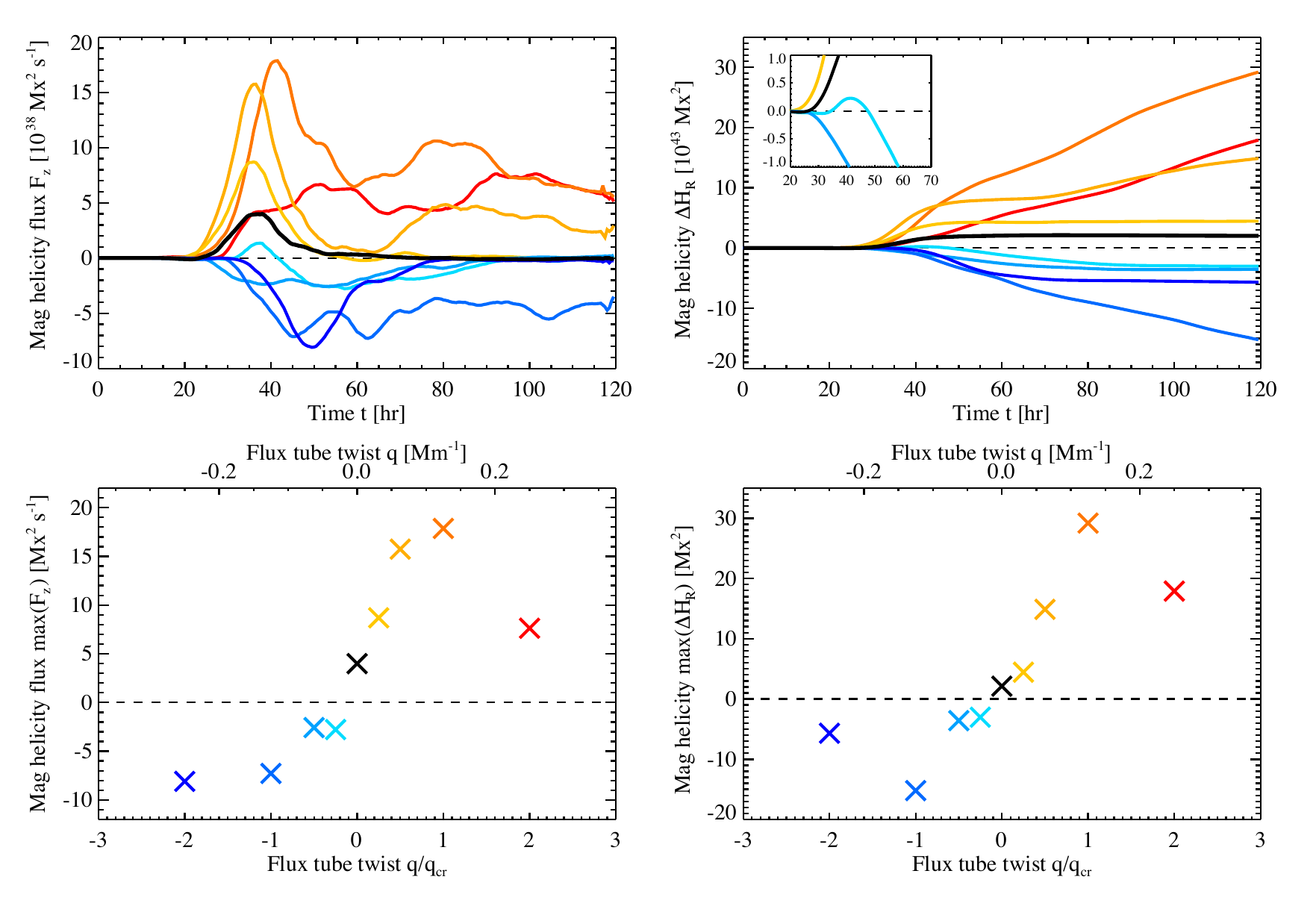}
\end{center}
\caption{(Top) Temporal evolutions of the helicity flux rate, $F_{z}$, and the total injected magnetic helicity, $\Delta H_{\rm R}$. The helicity $\Delta H_{\rm R}$ for the cases $q/q_{\rm cr}=[1/4, 0, -1/4, -1/2]$ are also shown in the inset. (Bottom) Their peak values as a function of the initial flux tube twist, $q/q_{\rm cr}$. For the negative twist cases ($q/q_{\rm cr}<0$), the peak values indicate the greatest negative values in the curves.\label{fig:helicity}}
\end{figure*}

The time variations of the helicity flux, $F_{z}$, and total injected helicity, $\Delta H_{\rm R}$, are shown in the top panels of Figure \ref{fig:helicity}. If we exclude the strongest twist case ($q/q_{\rm cr}=2$), the peak values reduce as the twist strength decreases, and $\Delta H_{\rm R}$ reaches $2.1\times 10^{43}\ {\rm Mx}^{2}\ {\rm s}^{-1}$ (that is, non-zero) for the twist-free case ($q/q_{\rm cr}=0$). Interestingly, for $q/q_{\rm cr}=-1/4$, both $F_{z}$ and $\Delta H_{\rm R}$ variations show a temporal excursion to the positive side at approximately $t=40\ {\rm hr}$, followed by an enhancement toward the negative side. This implies competition between the positive helicity added by the background convection and the counteracting negative magnetic helicity of the original flux tube.

The positive magnetic helicity owing to convection is clearly shown in the bottom panels of Figure \ref{fig:helicity}, which show the peak values of the helicity flux, $\max{(F_{z})}$, and injected helicity, $\max{(\Delta H_{\rm R})}$, for all nine cases. The values are biased toward the positive side in most cases, including $q/q_{\rm cr}=0$. However, in the strongest twist cases, for both positive and negative $q/q_{\rm cr}$, helicity injection appears to weaken, probably due to the reduced emerging magnetic flux (see Figure \ref{fig:phi}).

\begin{figure}
\begin{center}
\includegraphics[width=0.5\textwidth]{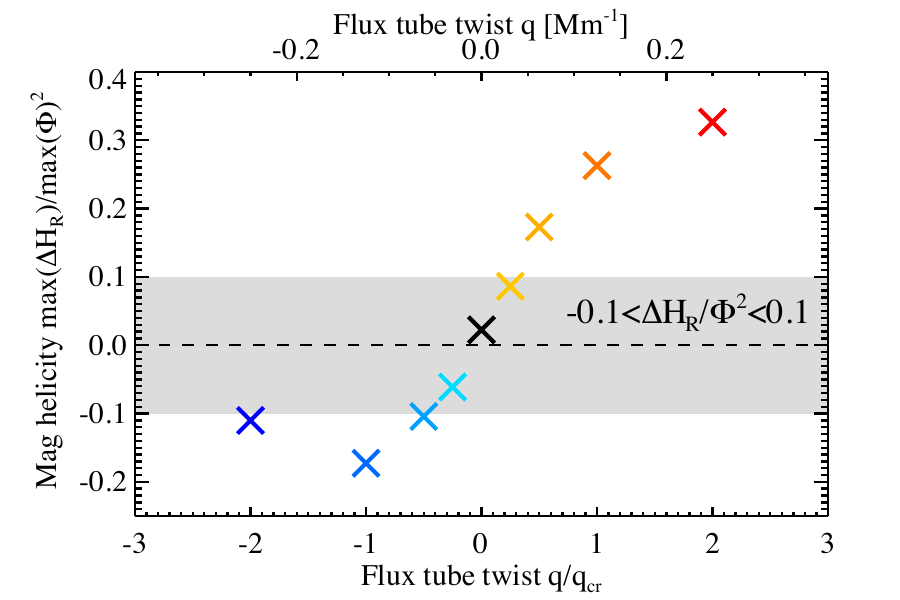}
\end{center}
\caption{Peak magnetic flux normalized by the square of the maximum magnetic flux, $\max{(\Delta H_{\rm R})}/\max{(\Phi)}^{2}$, as a function of the initial flux tube twist, $q/q_{\rm cr}$. The shaded area indicates the area where $-0.1<\Delta H_{\rm R}/\Phi^{2}<0.1$, which corresponds to the observational results.\label{fig:normhelicity_max}}
\end{figure}

In this study, we varied the twist of the initial flux tube over a broad parameter range from zero to exceeding the threshold for kink instability. To examine the validity of these calculations, we normalized the maximum magnetic helicity by the maximum magnetic flux, $\max{(\Delta H_{\rm R})}/\max{(\Phi)}^{2}$ \citep[see discussions in][]{2009AdSpR..43.1013D,2022arXiv220406010T}, and compared that with the observed values, which is summarized in Figure \ref{fig:normhelicity_max}.

Observations show that the typical value of the normalized helicity, $\Delta H_{\rm R}/\Phi^{2}$, is of the order of $0.01$ \citep{2007ApJ...671..955L,2024arXiv240318354S}, and even for the super-flaring ARs, the values are like $\lesssim 0.04$ (NOAA AR 11158), $\gtrsim 0.05$ (AR 12192), and $\sim 0.06$ (AR 12673) \citep{2019ApJ...887...64T,2019A&A...628A..50M}; that is, at most $0.1$ (indicated by the shade in Figure \ref{fig:normhelicity_max}). Our simulation shows that flux tubes with $|q/q_{\rm cr}|=1/2$ already yield a normalized helicity of $>0.1$; that is, they exceed the observed values for flaring ARs. Using even larger initial twists would result in a helicity injection far above the observed values. Therefore, setting the twist strength above the threshold for the kink instability ($|q/q_{\rm cr}|>1$) may not be realistic.

\subsection{Magnetic energy}\label{subsec:energy}

\begin{figure*}
\begin{center}
\includegraphics[width=0.9\textwidth]{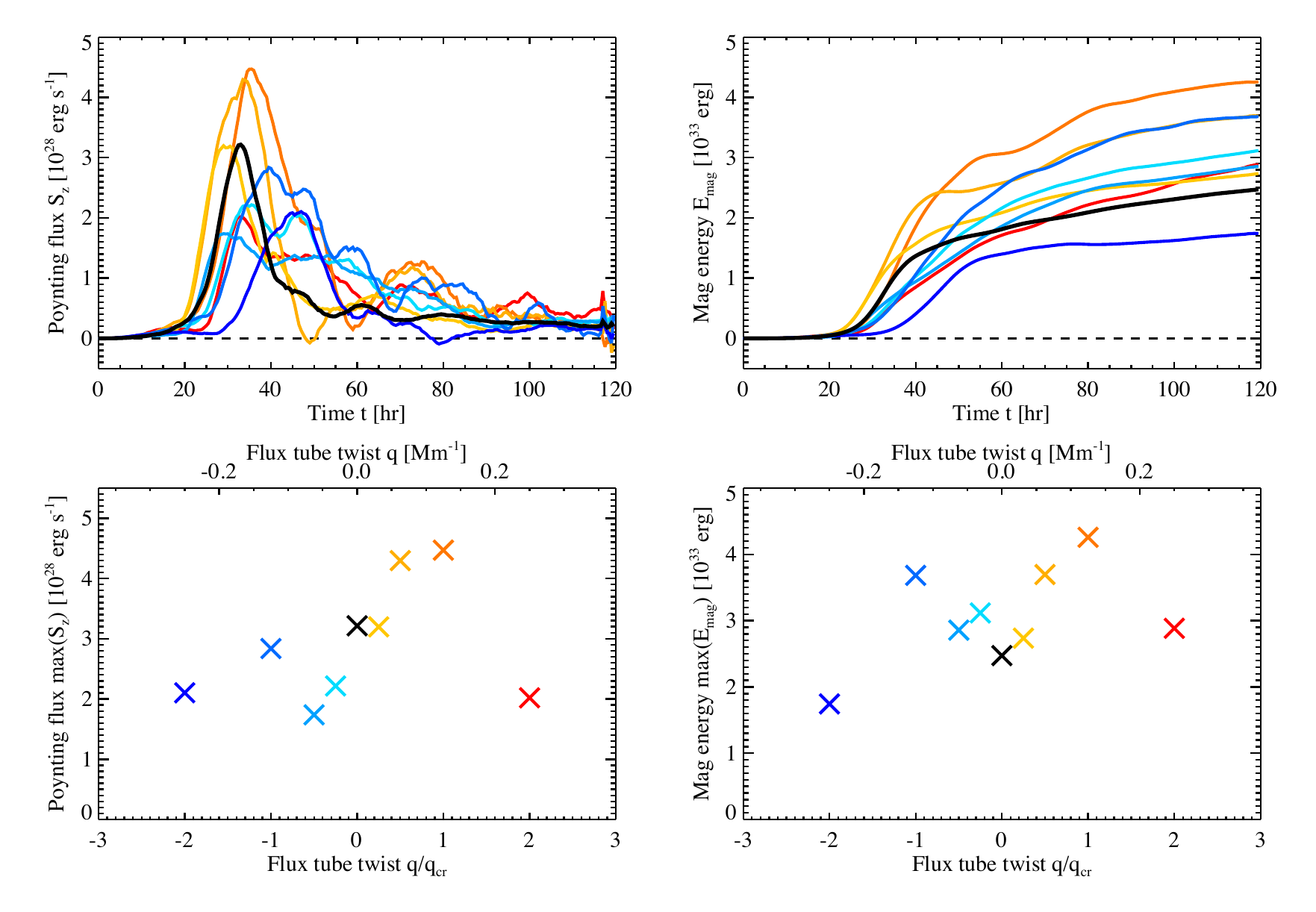}
\end{center}
\caption{(Top) Temporal evolutions of the Poynting flux in the photosphere, $S_{z}$, and the injected magnetic energy, $E_{\rm mag}$. (Bottom) Their peak values as a function of the initial flux tube twist, $q/q_{\rm cr}$.\label{fig:menergy}}
\end{figure*}

The temporal evolutions of the Poynting flux $S_{z}$, injected magnetic energy $E_{\rm mag}$, and their peak values are shown in Figure \ref{fig:menergy}. The Poynting flux shows time profiles similar to those of the flux growth rate, $d\Phi/dt$ (Figure \ref{fig:phi}), but with a higher variation between different twist cases. The peak values of the Poynting flux, $\max{(S_{z})}$, were highest for $q/q_{\rm cr}=1$ and $1/2$. However, interestingly, the $q/q_{\rm cr}=-1$ case shows a large peak value of the magnetic energy, $\max{(E_{\rm mag})}$, because of the continued injection of the Poynting flux, although its peak value, $\max{(S_{z})}$, is not necessarily large.

Overall, as the twist $q/q_{\rm cr}$ increases from $0$ in both the positive and negative directions, the net injected magnetic energy tends to increase, either because $\max{(S_{z})}$ is larger or because the emergence continues. However, when the twist is significantly larger at $q/q_{\rm cr}=\pm 2$, the magnetic energy is not remarkable because the flux emergence fails (Section \ref{subsec:flux}).

The amount of $E_{\rm mag}$ transported into the atmosphere is $(1.74$--$4.25)\times 10^{33}\ {\rm erg}$, which is only 3\%--7\% of the total magnetic energy of the original subphotospheric flux tube.

\section{Summary and Discussion}\label{sec:discussion}

In this study, we performed a series of convective flux emergence simulations in a deep domain using the R2D2 code to investigate the dependence of the flux tube on the initial twist strength, $q/q_{\rm cr}$. The numerical setup is mostly the same as that in \citet{2023NatSR..13.8994T} with a Gaussian-type flux tube embedded at a depth of $22\ {\rm Mm}$ in a deep background convection, except for a wide parameter range of the initial twist strength, $-2\leq q/q_{\rm cr}\leq 2$.

Flux emergence simulations have broad applications and are extensively used to study solar jets and flares \citep{2011LRSP....8....6S,2014LRSP...11....3C}. Recently, they have been used to provide the photospheric boundary conditions to test the data-driven AR modeling \citep{2017ApJ...838..113L,2020ApJ...890..103T,2020ApJ...903...11J,2023ApJ...949..118C,2023ApJ...952..136A,2024ApJS..270...30T}. Therefore, examining how {\it realistic} these realistic models are is now increasingly important. In this regard, we compared our simulation results with observations to provide constraints on subphotospheric flux tubes from a theoretical perspective.

In Figure \ref{fig:phi}, we find that neither the total unsigned magnetic flux $\Phi$ nor the flux growth rate $d\Phi/dt$ exhibit a significant dependence on the initial twist. Rather, when the twist was the strongest ($q/q_{\rm cr}=\pm 2$), both $\Phi$ and $d\Phi/dt$ exhibited smaller values. This may be because the magnetic energies were initially made equal, resulting in a relatively weak field strength for these cases, and they may have been disturbed by strong background convection. In the weak-twist cases, specifically for the untwisted one ($q/q_{\rm cr}=0$), the spot area $A_{\rm spot}$ tended to be smaller in relation to the total magnetic flux $\Phi$ (Figure \ref{fig:usflux_spotarea}). This may be because a flux tube with less twisting cannot hold the magnetic flux in the sunspot against flux dispersal by granulation.

By comparing $\Phi$ and $d\Phi/dt$ with the actual observations (Figure \ref{fig:phi_dphidt}), we found that although $d\Phi/dt$ is a few times larger than most of the actual ARs, it is still within the range for AR 12673, the most violent AR in solar cycle 24; thus, our models are fairly accurate.

The degree of twisting in a photospheric magnetic field is often quantified using the twist parameter $\alpha_{\rm av}$. In Figure \ref{fig:alpha}, although it was observed that the $\alpha_{\rm av}$ values measured in our simulations varied by a factor depending on the definition of $\alpha_{\rm av}$, the $\alpha_{\rm av}$ values agreed with those predicted from the initial flux tube's $q$ according to the theoretical relationship $\alpha=2q$. This indicates that the flux tubes conserved the original twists during emergence. The obtained photospheric $\alpha_{\rm av}$ values are within the range of those observed for the actual Sun. However, for larger twist cases, the $\alpha_{\rm av}$ values deviated from the $\alpha=2q$ line, possibly because flux emergence was unsuccessful or because the twist was stripped away before reaching the photosphere.

In \citet{2023NatSR..13.8994T}, for the untwisted flux tube ($q/q_{\rm cr}=0$), an injection of non-zero positive magnetic helicity ($\Delta H_{\rm R}$) was observed. In this study, we increased the number of simulation cases and found, in Figure \ref{fig:helicity}, that the $q/q_{\rm cr}=-1/4$ case showed a temporal excursion to the positive side and then increased to the negative side, suggesting that a conflict exists between the initially endowed negative twist and superposition of positive helicity by background turbulence. Overall, both $F_{z}$ and $\Delta H_{\rm R}$ values were biased toward the positive side.

The normalized magnetic helicity, $\Delta H_{\rm R}/\Phi^{2}$, exceeds the values measured in the actual ARs ($\lesssim 0.1$) already at $|q/q_{\rm cr}|=1/2$ (Figure \ref{fig:normhelicity_max}). Kink instability, which occurs for $|q/q_{\rm cr}|\ge 1$, has been proposed as a promising formation mechanism of $\delta$-spots, which is a key indicator of flare-productive ARs. However, our results suggest that kink instability is less likely to be the mechanism for the $\delta$-spots, because $|q/q_{\rm cr}|\ge 1$ yields unrealistically strong magnetic helicity injections in the photosphere. If this is true, other scenarios, such as the interaction of multiple flux tubes \citep{2007A&A...470..709M,2018ApJ...857...83J} and emergence of a multi-buoyant segment flux tube \citep{2014SoPh..289.3351T,2019A&A...630A.134S}, would need to be explored more in the future \citep[see also discussions in][]{2016ApJ...820L..11J,2019ApJ...871...67C,2022AdSpR..70.1549T,2022ApJ...938..117N}.

The magnetic energy transported into the upper atmosphere by the flux emergence was found to be at most less than 10\% of that of the initial flux tube. This means that even if the flux tube succeeds in bodily emergence, most of the magnetic energy remains in the convection zone.

We were unable to clearly confirm the development of kink instability for the $|q/q_{\rm cr}|\geq 1$ cases. Although it is possible that as the flux tube rises and expands, the critical value of the instability (i.e., $q_{\rm cr}=1/R_{\rm tb}$) changes, no clear indication was observed. One reason for this may be that the initial depth of the tubes was not deep enough to allow sufficient time for the instability to grow. Also, according to \citet{1996ApJ...469..954L}, the unstable wavenumber of the kink instability is limited to $-q-\Delta k/2<k<-q+\Delta k/2$, where $\Delta k=4qR_{\rm tb}(q^{2}-q_{\rm cr}^{2})^{1/2}/3.83$. Thus, the instability may not grow much for the marginal ($|q/q_{\rm cr}|\sim 1$) cases.

The simulations in this study were performed by placing the initial flux tubes at identical locations in the same background convection to focus on the dependence of the initial flux tube twist. However, as \citet{2022MNRAS.517.2775K} revealed that the initial location of the flux tube relative to the background convection has a significant impact on the fate of emergence, it is important to explore an even broader parameter space to better understand the nature of flux emergence and AR formation.

\begin{acknowledgments}
The authors wish to thank the anonymous referee for the comments and suggestions.
% Fugaku
The initial convection data were obtained using the Supercomputer Fugaku provided by the RIKEN Center for Computational Science.
% CfCA
Numerical computations were in part carried out on Cray XC50 at Center for Computational Astrophysics, National Astronomical Observatory of Japan.
% funding
This work was supported by JSPS KAKENHI Grant Nos. JP20KK0072 (PI: S. Toriumi), JP20K14510 (PI: H. Hotta), JP21H01124 (PI: T. Yokoyama), JP21H04492 (PI: K. Kusano), JP21H04497 (PI: H. Miyahara), and by
%MEXT as a Program for Promoting Researches on the Supercomputer Fugaku.
MEXT as ``Program for Promoting Researches on the Supercomputer Fugaku'' (JPMXP102023050).
\end{acknowledgments}


\vspace{5mm}
\facilities{RIKEN Supercomputer Fugaku, NAOJ CfCA Cray XC50}


\software{R2D2 \citep{2019SciA....5.2307H}}



\appendix

\section{Summary of the measured parameters}

Table \ref{tab:summary} summarizes the measured parameters for the nine simulation cases.

\begin{longrotatetable}
\begin{deluxetable*}{ccDDcDDDDDDDD}
\tablecaption{Summary of the measured parameters\label{tab:summary}}
\tablewidth{700pt}
\tabletypesize{\scriptsize}
\tablehead{
\colhead{Case} & \colhead{$q/q_{\rm cr}$} & \multicolumn2c{$\max{(\Phi)}$} & \multicolumn2c{$\max{(d\Phi/dt)}$} & \colhead{$A_{\rm spot}$} & \multicolumn2c{$\max(\alpha_{\rm av}^{(0)})$} & \multicolumn2c{$\max(\alpha_{\rm av}^{(1)})$} & \multicolumn2c{$\max(\alpha_{\rm av}^{(2)})$} & \multicolumn2c{$\max{(dH_{\rm R}/dt)}$} & \multicolumn2c{$\max{(H_{\rm R})}$} & \multicolumn2c{$\max{(H_{\rm R})}/\max{(\Phi)}^{2}$} & \multicolumn2c{$\max{(S_{z})}$} & \multicolumn2c{Case}\\
\nocolhead{} & \nocolhead{} & \multicolumn2c{(Mx)} & \multicolumn2c{(${\rm Mx\ hr}^{-1}$)} & \colhead{(MSH)} & \multicolumn2c{(${\rm Mx}^{-1}$)} & \multicolumn2c{(${\rm Mx}^{-1}$)} & \multicolumn2c{(${\rm Mx}^{-1}$)} & \multicolumn2c{(${\rm Mx}^{2}\ {\rm s}^{-1}$)} & \multicolumn2c{(${\rm Mx}^{2}$)} & \multicolumn2c{} & \multicolumn2c{(${\rm erg\ s}^{-1}$)} & \multicolumn2c{(erg)}
}
\decimalcolnumbers
\startdata
1 & $-2$ & $2.27\times 10^{22}$ & $1.26\times 10^{21}$ & $235$ & $-0.291$ & $-0.165$ & $-0.082$ & $-8.09\times 10^{38}$ & $-5.69\times 10^{43}$ & $-0.110$ & $2.10\times 10^{28}$ & $1.74\times 10^{33}$ \\
2 & $-1$ & $2.97\times 10^{22}$ & $1.47\times 10^{21}$ & $354$ & $-0.219$ & $-0.118$ & $-0.071$ & $-7.28\times 10^{38}$ & $-15.21\times 10^{43}$ & $-0.173$ & $2.84\times 10^{28}$ & $3.68\times 10^{33}$ \\
3 & $-1/2$ & $1.86\times 10^{22}$ & $1.38\times 10^{21}$ & $104$ & $-0.214$ & $-0.150$ & $-0.101$ & $-2.57\times 10^{38}$ & $-3.59\times 10^{43}$ & $-0.104$ & $1.74\times 10^{28}$ & $2.86\times 10^{33}$ \\
4 & $-1/4$ & $2.22\times 10^{22}$ & $1.39\times 10^{21}$ & $237$ & $-0.181$ & $-0.114$ & $-0.087$ & $-2.77\times 10^{38}$ & $-3.02\times 10^{43}$ & $-0.061$ & $2.22\times 10^{28}$ & $3.12\times 10^{33}$ \\
5 & $0$ & $3.07\times 10^{22}$ & $1.85\times 10^{21}$ & $221$ & $0.137$ & $0.097$ & $0.075$ & $3.99\times 10^{38}$ & $2.12\times 10^{43}$ & $0.022$ & $3.21\times 10^{28}$ & $2.47\times 10^{33}$ \\
6 & $1/4$ & $2.22\times 10^{22}$ & $1.39\times 10^{21}$ & $237$ & $0.032$ & $0.039$ & $0.037$ & $1.35\times 10^{38}$ & $0.23\times 10^{43}$ & $0.005$ & $2.22\times 10^{28}$ & $3.12\times 10^{33}$ \\
7 & $1/2$ & $2.27\times 10^{22}$ & $1.88\times 10^{21}$ & $213$ & $0.200$ & $0.162$ & $0.125$ & $8.69\times 10^{38}$ & $4.45\times 10^{43}$ & $0.086$ & $3.20\times 10^{28}$ & $2.74\times 10^{33}$ \\
8 & $1$ & $3.33\times 10^{22}$ & $2.15\times 10^{21}$ & $403$ & $0.308$ & $0.172$ & $0.101$ & $17.86\times 10^{38}$ & $29.19\times 10^{43}$ & $0.263$ & $4.47\times 10^{28}$ & $4.26\times 10^{33}$ \\
9 & $2$ & $2.34\times 10^{22}$ & $1.55\times 10^{21}$ & $250$ & $0.286$ & $0.150$ & $0.089$ & $7.63\times 10^{38}$ & $17.91\times 10^{43}$ & $0.327$ & $2.02\times 10^{28}$ & $2.88\times 10^{33}$ \\
\enddata
\tablecomments{$A_{\rm spot}$ in column (5) indicates the sunspot area at the time of $\max{(\Phi)}$. $\alpha_{\rm av}$ in columns (6), (7), and (8) are measured in the time period from $t=10\ {\rm hr}$ to $110\ {\rm hr}$.}
\end{deluxetable*}
\end{longrotatetable}


\bibliography{toriumi2024}{}
\bibliographystyle{aasjournal}



\end{document}